\documentclass[a4paper,12pt]{article}
\usepackage[hmargin=2cm,vmargin=2cm]{geometry}
\usepackage{amsmath,amssymb}
\usepackage{graphicx}
\usepackage[colorlinks=true, allcolors=blue]{hyperref}
\usepackage{physics}
\usepackage{amsmath,amsfonts,amssymb,amsthm,mathtools}
\usepackage{cmap} 
\usepackage{braket}
\usepackage{authblk}

\usepackage{comment}

\newcommand{\sign}{\text{sign}}
\newcommand{\Br}{\text{Br}}
\newcommand{\asymm}{\text{asymm}}
\newcommand{\symm}{\text{symm}}
\newcommand{\mix}{\text{mix}}
\newcommand{\peng}{\text{peng}}

\title{Correction to $CP$-asymmetry in $\Upsilon(4S)$ decays\\
due to the admixture of $B^0 \bar B^0$ in a $C$-even state}
\author[1]{N.A. Panchenko}
\author[2]{S.I. Godunov}
\affil[1]{\textit{Moscow Institute of Physics and Technology, Dolgoprudny 141700, Russia}}
\affil[2]{\textit{I.E. Tamm Department of Theoretical Physics, Lebedev Physical Institute, 53 Leninskiy Prospekt, Moscow, 119991, Russia}}

\date{}

\begin{document}

\maketitle

\begin{abstract}
$BB$ pairs from $\Upsilon(4S)\to BB$ decays are in $C$-odd state. However, there is a small admixture of the $C$-even state and it modifies the time dependent $CP$-asymmetry. The $C$-even component appears due to soft photon emission, breaking the pure $C$-odd nature of the initial state. Using the two-particle wave function formalism, we derive analytical expressions for $CP$-asymmetry in both pure and mixed $C$-parity states. We emphasize the strong energy dependence of the admixture magnitude.
\end{abstract}

\section{Introduction}

In the decay of the vector $\Upsilon(4S)$-meson into two pseudoscalar $B^0$-mesons, the $B^0\bar{B}^0$ pair is produced in a P-wave. This follows from the law of angular momentum conservation: the total spin of the pair $S=0$, consequently $J = L = 1$. Now one can determine the $P$- and $C$-parities of the system: $C = P = (-1)^L = -1$. Thus, the standard wave function of the $B^0\bar{B}^0$ pair produced in the $\Upsilon(4S)$ decay is $C$-odd and antisymmetric with respect to meson interchange. The emission of a photon in this decay changes the state to $C$-even. We investigate how this affects the time-dependent $CP$-asymmetry.

The motivation for this study is the search for and quantitative assessment of fundamental bounds on the accuracy of measuring the angles of the unitarity triangle from time-dependent $CP$-asymmetry in $B$-meson decays. The parameters of the unitarity triangle are extracted from the $CP$-asymmetry \cite{Belle-II:2018jsg}. Various processes contribute to the total measurement error, some of which impose a fundamental limitation that cannot be eliminated by increasing statistics. In this work, we focus on one such process – the emission of a soft photon in the $\Upsilon(4S)$ decay, which, as described above, leads to the formation of a $B^0\bar{B}^0$ pair in a $C$-even state.

To address this issue, we considered a process in which one $B^0$-meson of the pair decays semileptonically into the final state $\ell^{\pm}\nu X$, and the second into a $CP$-eigenstate $f$, when the system is in a $C$-even and $C$-odd state. For both cases, the number of decay events into a lepton of negative charge $N_- \equiv N(B^0 \bar B^0 \rightarrow \ell^- \bar \nu X,f)$ and via the $CP$-conjugate channel $N_+ \equiv N(B^0 \bar B^0 \rightarrow \ell^+ \nu X, \bar f)$ was calculated. Using this data, we found the $CP$-asymmetry, taking into account the admixture of the decay probability from the state with the symmetric wave function of the pair to the decay probability from the state with the antisymmetric one.

\section{Two-particle wave function formalism}
If we know the $B$-meson flavor at some moment, then its decay can be described within the framework of a single-particle problem. When the pair is in a $C$-odd state, this is possible thanks to so-called ``tagging''. But this approach is not applicable in cases where the pair is in the C-even state. Then, the pair should be described using a two-particle wave function. Thus, in this section we consider both cases in the two-particle framework.

\subsection{$C$-odd state of the $B^0\bar{B}^0$ pair}
As noted earlier, when the wave function of the $B^0$-mesons system is antisymmetric under permutations, we have:
\begin{equation} 
      \Psi_{\asymm}(t) = B^{0}_1(t)\bar B^{0}_2(t) - \bar B^{0}_1(t) B^{0}_2(t),
\end{equation}
where $B^0(t)$ and $\bar B^0(t)$ are functions describing the time evolution of neutral $B$-mesons:
\begin{equation}
    B^{0}(t) = e^{-i\frac{(M_{H} + M_{L})t}{2} - \frac{\Gamma t}{2}}
\left[\cos{\left(\dfrac{\Delta m t }{2}\right)} B^{0} + i \dfrac{q}{p} \sin{\left(\dfrac{\Delta m t }{2}\right)} \bar B^{0}\right],
\end{equation}
\begin{equation} 
    \bar B^{0}(t) = e^{-i\frac{(M_{H} + M_{L})t}{2} - \frac{\Gamma t}{2}}
\left[i \dfrac{p}{q} \sin{\left(\dfrac{\Delta m t }{2}\right)} B^{0} + \cos{\left(\dfrac{\Delta m t }{2}\right)} \bar B^{0} \right],
\end{equation}
where $p$ and $q$ are the parameters describing the mixing of $B$-mesons, $M_H$ and $M_L$ are the masses of $B_H$ and $B_L$, which are the mass eigenstates of the system, $\Delta m=M_H - M_L$, $\Gamma = \Gamma_H = \Gamma_L$ is their width (we neglect the difference between $\Gamma_H$ and $\Gamma_L$ for our purposes). Here and everywhere below, unless otherwise stated, we use the notations from \cite{ParticleDataGroup:2024cfk}.

The decay amplitude is obtained by projecting the final state onto this wave function:
\begin{equation} \label{A_asymm}
     \braket{\ell^{-}(t_1),f(t_2)|\Psi_{\asymm}(t)} = -e^{-2i\left(m - i\frac{\Gamma}{2}\right)t}A_- \left( A_f \cos{\dfrac{\Delta t \Delta m}{2}} + i\dfrac{q}{p} \bar A_f \sin{\dfrac{\Delta t \Delta m}{2}}\right),
\end{equation}
where $t_1$ and $t_2$ are the moments in time when $B^0_1(t)$ and $B^0_2(t)$ decayed respectively, $t \equiv \frac{t_2 + t_1}{2}$, $\Delta t \equiv t_2 - t_1$, $m=(M_H+M_L)/2$, $A_f$ denote the decay amplitude into the $CP$-eigenstate $f$, $A_-$ denotes the semileptonic decay amplitude.

To obtain the number of events, we need to integrate the decay probability over time. Let's replace the integration limits:
\begin{equation}
    \int_{0}^{\infty} d t_1 \int_{0}^{\infty} dt_2 = \int_{-\infty}^{+\infty} d(\Delta t) \int_{\frac{\abs{\Delta t}}{2}}^{+\infty} dt.
\end{equation} 
It is important to note that the lower limit of integration over $t$ is the absolute value of the difference in decay times. As one can see from \eqref{A_asymm}, the entire dependence on $t$ is contained solely in the power of the exponent; sine and cosine depend only on $\Delta t$. Thus, during integration, the $\abs{\Delta t}$ will appear only in the power of the exponent. For the event rate we get:
\begin{equation} 
    N_-^{\asymm}(\Delta t) \propto \dfrac{1}{2\Gamma}e^{-\Gamma \abs{\Delta t}}\abs{A_-}^2\abs{A_f}^2 \left[ \cos^2{\dfrac{\Delta t\Delta m}{2}} +\abs{\lambda}^2 \sin^2{\dfrac{\Delta t\Delta m}{2}} - \Im\lambda \sin{\Delta t \Delta m}\right],
\end{equation}
where $\lambda = \dfrac{q \bar A_f}{p A_f}$.

We apply the $CP$-conjugation operator to this process and similarly obtain an expression for the number of events with a positive lepton in the final state:
\begin{equation}  \label{NCPsopr_asym}
    N_+^{\asymm}(\Delta t) \propto \dfrac{1}{2\Gamma}e^{-\Gamma \abs{\Delta t}}\abs{A_+}^2\abs{\bar A_f}^2 \left[ \cos^2{\dfrac{\Delta t\Delta m}{2}} +\abs{\frac{1}{\lambda}}^2 \sin^2{\dfrac{\Delta t\Delta m}{2}} - \Im \frac{1 }{\lambda} \sin{\Delta t \Delta m}\right]. 
\end{equation}
Thus, the standard expression for $CP$-asymmetry \cite{ParticleDataGroup:2024cfk} is obtained
\begin{equation}
    \label{CPasgeneral}
    a_{CP}^{\asymm}(\Delta t) = \dfrac{N_+^{\asymm}(\Delta t) - N_-^{\asymm}(\Delta t)}{N_+^{\asymm}(\Delta t) + N_-^{\asymm}(\Delta t)} \equiv - C_f \cos{\Delta t \Delta m} + S_f \sin{\Delta t \Delta m},
\end{equation}
where
\begin{equation}
    C_f = \dfrac{1 - \abs{\lambda}^2}{1+\abs{\lambda}^2}, \qquad S_f = \dfrac{2\Im \lambda}{1 + \abs{\lambda}^2}.
\end{equation}

\subsection{$C$-even state of the $B^0\bar{B}^0$ pair}
In the case where the pair is in the $C$-even state, its wave function takes the following form
\begin{equation}
    \Psi_{\symm}(t) = B^0_1(t)\bar B^0_2(t) + B^0_2(t) \bar B^0_1(t).
\end{equation}
Similar to the case considered earlier, we find the meson decay amplitude when their wave function is symmetric under permutations
\begin{equation}
     \braket{\ell^{-}(t_1),f(t_2)|\Psi_{\symm}} = e^{-2i\left(m - i\frac{\Gamma}{2}\right)t}A_-\left(A_f \cos{t\Delta m} + i\dfrac{q}{p} \bar A_f \sin{{t\Delta m}}\right).
\end{equation}
Here, it should be noted that, unlike the case where the pair is in the $C$-odd state, the dependence on $t$ is contained not only in the power of the exponent but also in the arguments of the sine and cosine. Thus, when we integrate the probability over time, that is, find the number of events, $\abs{\Delta t}$ will also appear in the arguments of the sine and cosine. This is a very important difference between the $C$-even and $C$-odd states.

Number of events in two $CP$-conjugate processes and $CP$-asymmetry:
\begin{align}
     N_-^{\symm} &\propto \dfrac{\abs{A_-}^2 \abs{A_f}^2 e^{-\Gamma \abs{\Delta t}}}{4 \Gamma}\Bigg[\left(1 + \abs{\lambda}^2\right)  -2\Im \lambda \dfrac{x \cos{\abs{\Delta t}\Delta m} + \sin{\abs{\Delta t} \Delta m} }{1+ x^2} + \nonumber\\ 
     &
     \hphantom{\propto \dfrac{\abs{A_-}^2 \abs{A_f}^2 e^{-\Gamma \abs{\Delta t}}}{4 \Gamma}\Bigg[}+ \left(1 - \abs{\lambda}^2\right)\dfrac{\cos{\abs{\Delta t}\Delta m} -x\sin{\abs{\Delta t} \Delta m}}{1+ x^2} 
     \Bigg],\\
    N_+^{\symm} &\propto \dfrac{\abs{A_+}^2 \abs{\bar A_f}^2 e^{-\Gamma \abs{\Delta t}}}{4\Gamma}\Bigg[\left(1+ \dfrac{1}{\abs{\lambda}^2}\right) -2\Im \dfrac{1}{\lambda} \dfrac{x \cos{\abs{\Delta t}\Delta m} + \sin{\abs{\Delta t} \Delta m} }{1+ x^2} + \nonumber \\
    &
    \hphantom{\propto \dfrac{\abs{A_+}^2 \abs{\bar A_f}^2 e^{-\Gamma \abs{\Delta t}}}{4\Gamma}\Bigg[}+ \left(1 - \dfrac{1}{\abs{\lambda}^2}\right)\dfrac{\cos{\abs{\Delta t}\Delta m} -x\sin{\abs{\Delta t} \Delta m}}{1+ x^2}  \Bigg],\\
    a_{CP}^{\symm}(\Delta t) &= \dfrac{1}{1+\abs{\lambda}^2} \dfrac{1}{1+ x^2}\Bigg[ -\left[\left(1 - \abs{\lambda}^2 \right) - 2x\Im\lambda \right]\cos{\Delta t\Delta m} + \nonumber \\
    &
    \hphantom{= \dfrac{1}{1+\abs{\lambda}^2} \dfrac{1}{1+ x^2}\Bigg[}+ \left[\left(1 - \abs{\lambda}^2 \right)x + 2 \Im\lambda \right]\sign{\Delta t}\sin{\Delta t\Delta m}\Bigg],
    \label{CPasym_sym}
\end{align}
where $x = \frac{\Delta m}{\Gamma}$. Dependence on the sign of $\Delta t$ in the expression \eqref{CPasym_sym} is different from what we had in~\eqref{CPasgeneral}: in the case of a $C$-even state of a $B^0$ mesons pair, the $CP$-asymmetry does not depend on the order of decays, while in the case of a $C$-odd state, it does.

\section{$CP$-asymmetry for a state with mixed $C$-parity}
A pair of $B^0$-mesons produced in the decay of a vector $\Upsilon(4S)$-meson has an antisymmetric wave function. The emission of a photon leads to a symmetric wave function. The final states in these two cases are different: $B^0 \bar B^0$ and $B^0 \bar B^0 \gamma$, respectively. Therefore, there is no interference between the two asymmetries:
\begin{equation} \label{mix acp}
\begin{aligned}
   a^{\mix}_{CP}(\Delta t) &= \dfrac{a^{\asymm}_{CP}(\Delta t)}{1+\chi} + \dfrac{\chi a^{\symm}_{CP}(\Delta t)}{1+\chi} ,
\end{aligned}
\end{equation}
where $\chi$ is the admixture value (relative to the $C$-odd rate). Let us consider expression \eqref{mix acp} in more detail using a specific decay $B^0 \to J/\psi K_{S(L)}$ --- the ``gold-plated mode'', in which the $\beta$ angle of the unitarity triangle was measured by the Belle \cite{Belle:2012paq} and BaBar \cite{BaBar:2001pki} collaborations. Then $\abs{\lambda} = 1$ with good accuracy and $\Im \lambda = - \eta_f \sin{2\beta}$:
\begin{equation} \label{cp mix}
   a^{\mix}_{CP}(\Delta t) =- \frac{\eta_f \sin{2\beta}}{1 + \chi}\left[\left(1 + \sign(\Delta t)\frac{\chi}{1+x^2}\right)\sin{\Delta t \Delta m} + \chi \frac{x}{1+x^2}\cos{\Delta t \Delta m}\right],
\end{equation}
where $\eta_f = \pm 1$ ($\eta_S = - 1$, $\eta_L = +1$) is the eigenvalue of the $CP$-conjugation operator.

Recalling the expression for the $CP$-asymmetry in the case of the antisymmetric wave function of $B^0 \bar B^0$-mesons \eqref{CPasgeneral}, it can be shown that the corrections to the coefficients of the sine and cosine have the form:
\begin{equation}\label{correction sin cos}
    \Delta C^{\mix} = C^{\mix} =\frac{\chi}{1 + \chi} \dfrac{x}{1+x^2}\eta_f\sin{2\beta}, \qquad
    \Delta S^{\mix} = \frac{\chi}{1 + \chi} \left[1 - \dfrac{\sign{\Delta t}}{1 + x^2}\right] \eta_f\sin{2\beta}.
\end{equation} 

\begin{figure}[t]
    \centering
    \includegraphics[width=6.48in]{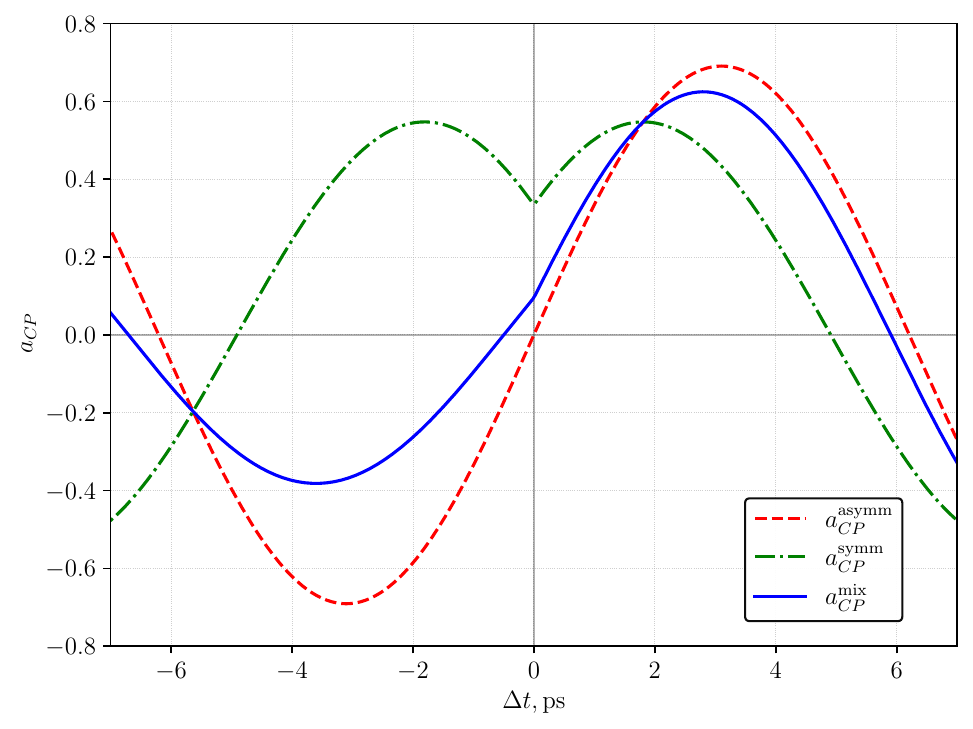}
    \caption{Time-dependent $CP$-asymmetry in the case of $C$-odd (red dashed) and $C$-even (green dot-dashed) states and their mixture (blue solid) with an admixture of $\chi = 0.4$ in the case of $\eta_S = -1$.}
    \label{fig:myfigure}
\end{figure}

Fig.~\ref{fig:myfigure} shows the $CP$-asymmetry as a function of $\Delta t$ for three configurations: pure $C$-odd state (red dashed), pure $C$-even state (green dot-dashed), and mixed state (blue solid) at $\chi$=0.4, with parameter $\eta_S = -1$. The reason for choosing this value is the decays of $\Upsilon(5S)$, in which a pair of $B^0 \bar B^0$-mesons with a $C$-even wave function is produced with a probability of $11\%$, and with a $C$-odd one --- $27\%$ \cite{bondar}. Thus, the choice of the parameter $\chi$ = 0.4 is not only a convenient and illustrative example for our calculations but also demonstrates how the $CP$-asymmetry changes when the influence of the $C$-even pair is taken into account in the case of the $\Upsilon(5S)$ decay.

The introduction of a $C$-even admixture produces several characteristic modifications of the time-dependent $CP$-asymmetry. In contrast to the purely $C$-odd case, the asymmetry does not vanish at $\Delta t = 0$. Moreover, the curve exhibits suppression, which is stronger at $\Delta t < 0 $. Finally, the zero-crossing point of the asymmetry is shifted toward negative values of $\Delta t$.

\section{The value of the correction}

In the paper~\cite{colangelo} a mechanism for the formation of the $B^0 \bar B^0$ pair through virtual $B^*$ is studied
\begin{equation} \label{channel}
    \Upsilon(4S) \to B^* \bar B^0 \to B^0 \bar B^0 \gamma.
\end{equation}
As the authors note, this mechanism is potentially important due to the small difference between the masses of $B^*$ and $B^0$. 
The branching of the $\Upsilon(4S) \rightarrow B^* \bar B^0 \rightarrow B^0 \bar B^0\gamma$ channel was calculated in~\cite{colangelo}, giving the admixture of the $C$-even state. We have reproduced this result to investigate its sensitivity to the center-of-mass energy. Specifically, we will vary the center-of-mass energy $\sqrt{s}$ around the $\Upsilon(4S)$ mass. As it will be demonstrated, moving off the resonance peak results in a strong enhancement of the branching fraction.

The decay width of the process \eqref{channel}:
\begin{equation}
   \label{width3_3}
   \Gamma = \dfrac{\lambda^2 g^2_{B^0 B^*}}{9 \pi^3} \int_{M_{B^0}}^{\frac{M_{\Upsilon}}{2}} \dfrac{M_{\Upsilon}^3 M_{B}^2 \left(E_3^2 - M_{B}^2\right)^{\frac{3}{2}}\left(M_{\Upsilon} - 2E_3\right)^3 dE_3}{\left(M_{\Upsilon}^2 + M_{B}^2 - 2 E_3 M_{\Upsilon}\right)\left(\left(M_{\Upsilon}^2 + M_{B}^2 - 2M_{\Upsilon}E_3- M_{B^*}^2\right)^2 + M_{B^*}^2 \Gamma_{B^*}^2\right)},
\end{equation}
where $g_{B^0 B^*}$ and $\lambda$ are the coupling constants at the interaction vertices $B^* \to B^0 \gamma$ and $\Upsilon(4S) \to B^* \bar B^0$, respectively.

The paper \cite{colangelo} presents the results:
\begin{equation}
    \Br(\Upsilon(4S) \to B^0 \bar B^0 \gamma) = 3 \cdot 10^{-9}.
\end{equation}
It is interesting that, using modern values of particle masses, we get $\Br(\Upsilon(4S) \to B^0 \bar B^0 \gamma) \simeq 2 \cdot 10^{-9}$ from~\eqref{width3_3}, which is 1.5 times lower than the original estimate. It is noteworthy that even a relatively small shift of the masses results in a sizable change in the predicted branching ratio. According to \eqref{correction sin cos}, the corrections to the coefficients of sine and cosine are

\begin{equation}
    \Delta C^{\mix} = 1.3 \cdot 10^{-9}\eta_f , \qquad
    \Delta S^{\mix} = 
  \begin{cases} 
   10^{-9}\eta_f & \text{, when } \Delta t > 0 \\
   4.5 \cdot 10^{-9}\eta_{f} & \text{, when } \Delta t < 0
  \end{cases}.
\end{equation}
  
\begin{figure}[t] 
    \centering
    \includegraphics[width=6.11in]{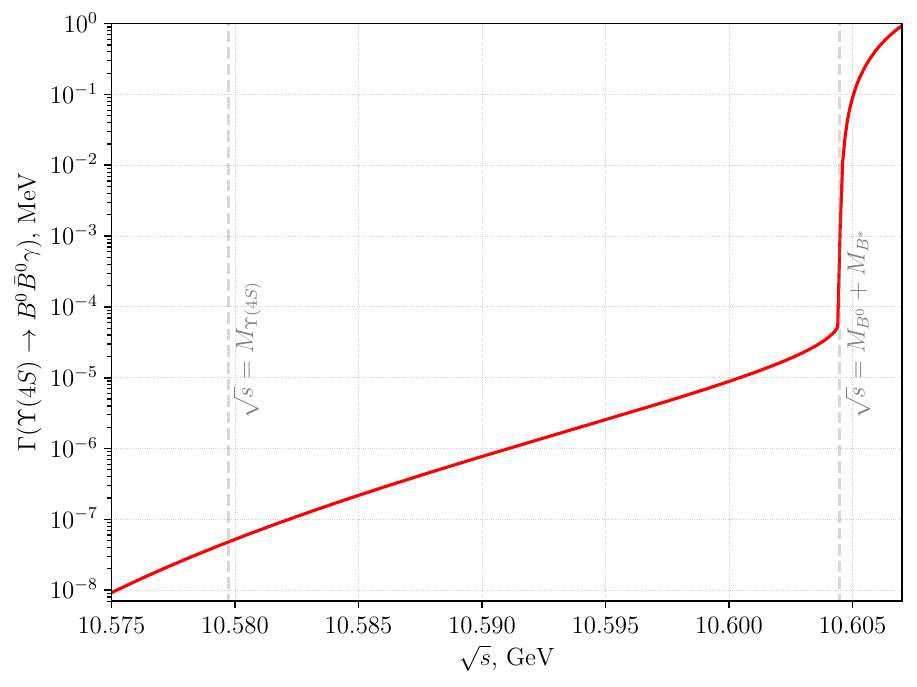}
    \caption{The decay width of $\Upsilon(4S) \to B^0 \bar B^0 \gamma$ as a function of $\sqrt{s}$.}
    \label{brench}
\end{figure}

The branching ratio of the $\Upsilon(4S)\to B^0 \bar B^0 \gamma$ channel could be further increased by varying $\sqrt{s}$. The growth of the corresponding width is shown in Fig.~\ref{brench}. We see that the width rapidly (exponentially) grows with $\sqrt{s}$, and at the threshold of $B B^*$ production it exhibits a step-like behavior. It originates from the transition of the $B^*$ from a virtual to an on-shell state, which dramatically enhances the probability of the $\Upsilon(4S) \to B^0 B^*$ decay channel. Let us note that the width gains 3 orders of magnitude even before reaching $\Upsilon(4S)\to BB^*$ threshold.

Thus, an increase of the $\sqrt{s}$ by 25 MeV (i.e., within 1 -- 2 $\Upsilon(4S)$ widths) leads to a substantial enhancement of the radiative decay rate, reaching $\Gamma(\Upsilon(4S) \to B^0 \bar B^0 \gamma) \sim 0.1 $ MeV, corresponding to a branching fraction of the order of $\Br(\Upsilon(4S) \to B^0 \bar B^0 \gamma) \sim 10^{-3}$. Substituting this value into~\eqref{correction sin cos}, one finds that the corrections to the sine and cosine coefficients from the $C$-even admixture become $\Delta C^{\mix} \sim 10^{-4}$, $\Delta S^{\mix}(\Delta t < 0) \sim 10^{-3}$ and $\Delta S^{\mix}(\Delta t > 0) \sim 10^{-4}$. Therefore, if experimental data taking is performed away from the exact $\Upsilon(4S)$ peak, this $C$–even admixture effect would become non-negligible and must be explicitly accounted for in the extraction of the time–dependent $CP$–asymmetry coefficients.

There is another potential mechanism: virtual $\chi_{b0}(3P)$ and $\chi_{b2}(3P)$  will contribute via $\Upsilon(4S) \to \chi(3P) \gamma \to B^0 \bar B^0 \gamma$. A rough estimate of the branching ratio shows that it is no higher in order than the mechanism via the virtual $B^*$-meson that we considered. Moreover, these contributions do not gain so much from $\sqrt{s}$ increment.

\section{Penguin diagrams' contribution} 

In studying the $B^0 \rightarrow J/\psi K_{S(L)}$ decay, only the lowest-order perturbation diagram (tree diagrams) was considered in~\eqref{CPasgeneral}. However, taking into account the next orders can make a significant contribution to the $CP$-asymmetry. Adding the penguin amplitude will affect the time-dependent $CP$-asymmetry (so-called ``penguin pollution''), and according to \cite{peng}:
\begin{equation} \label{peng acp}
    a_{CP} = - C^{\peng} \cos{\Delta t\Delta m} + S^{\peng} \sin{\Delta t\Delta m},
\end{equation}
where are the coefficients of sine and cosine:
\begin{equation}
    \Delta C^{\peng} = C^{\peng} = - 2 \frac{P}{T} \sin{\delta} \sin{\gamma},
\end{equation}
\begin{equation}
    \Delta S^{\peng} \equiv S^{\peng} - \sin{2\beta} = 2 \frac{P}{T} \cos{2\beta} \cos{\delta} \sin{\gamma},
\end{equation}
where $\frac{P}{T}$ is the ratio of the amplitudes of the penguin diagram to the tree diagram, $\delta$ and $\gamma$ are the strong and weak phases, respectively.

According to \cite{pengValue}:
\begin{equation}  
\Delta S^{\peng}= \left(7.2^{+2.4(+1.2)}_{-3.4(-1.1)}\right) \cdot 10^{-4}, \qquad C^{\peng} = - \left(16.7^{+6.6(+3.8)}_{-8.7(-4.1)}\right) \cdot 10^{-4},
\end{equation}
which is 6 orders of magnitude greater than the corrections to the corresponding values from the contribution of the $C$-even wave function, if we assume that $\Upsilon(4S)$ is produced on-shell. However, if we increase the energy by 25 MeV, the correction from the $C$-even state becomes comparable to the penguin contribution.

\section{Conclusion}

In this work, we have analyzed the correction to the time-dependent $CP$-asymmetry in the decays of the $B^0\bar{B}^0$ pair from the $\Upsilon(4S)$-resonance caused by the admixture of a $C$-even component in the two-meson wave function. Such a component appears as a result of soft photon emission, which changes the $C$-parity of the system and makes the state symmetric under particle exchange.

We have obtained analytical expressions for the time-dependent $CP$-asymmetry in the case of a purely $C$-even state as well as for the mixed state of the $B^0\bar{B}^0$ pair. An important feature of the $C$-even case is that the $CP$ asymmetry depends on the absolute value of the decay time difference, i.e., on $|\Delta t|$, unlike the $C$-odd case. The behavior of the asymmetry in the presence of a $C$-even admixture turns out to be qualitatively different compared to the well-known case of a purely $C$-odd state: the asymmetry is suppressed, in $\Delta t < 0$ region this suppression is stronger, and the position of the zero crossing of the asymmetry is shifted to the left. These effects are illustrated using the ``gold-plated mode'' $B^0 \to J/\psi\,K_S$ \eqref{cp mix}.

A quantitative estimate shows that the correction due to the $C$-even state is extremely small for $\sqrt{s} = M_{\Upsilon(4S)}$, corresponding to $\Br(\Upsilon(4S)\to B^0\bar{B}^0\gamma)\sim 10^{-9}$. However, with only a slight upward shift from the resonance mass (by about 25 MeV), the correction can reach $\mathcal{O}(10^{-4})$, making it comparable to the well-known “penguin pollution” effects. 

\section*{Acknowledgement}
We sincerely thank A.~E.~Bondar and M.~I.~Vysotsky for their guidance and enlightening discussions. We are also grateful to P.~N.~Pakhlov, T.~V.~Uglov and M.~I.~Yasaveev for the valuable discussions and ideas.

\bibliographystyle{unsrturl}
\bibliography{references}

\end{document}